\documentclass[twocolumn,showpacs,amsmath,amssymb,prl]{revtex4}
\usepackage{dcolumn}% Align table columns on decimal point
\usepackage{epsfig}
\usepackage[latin1]{inputenc}
\usepackage{bm}% bold math
\begin{document}

%\documentstyle[eqsecnum,multicol,epsfig,aps,prb,array]{revtex}
%\newcommand{\bleq}{\ifpreprintsty
%		   \else
%		   \end{multicols}\widetext \vspace*{-3.5ex}{\tiny
%		
%		\noindent\begin{tabular}[t]{c|}
%		   \parbox{0.493\hsize}{~} \\ \hline \end{tabular}}
%				      \fi}
%\newcommand{\eleq}{\ifpreprintsty
%		   \else
%		   {\tiny\hspace*{\fill}\begin{tabular}[t]{|c}\hline
%		    \parbox{0.49\hsize}{~} \\
%		    \end{tabular}}\vspace*{-2.5ex}\begin{multicols}{2}
%		    \narrowtext
%		    \fi}
%\newcommand{\bcols}{\ifpreprintsty\else\begin{multicols}{2}
%	\narrowtext\fi}
%\newcommand{\ecols}{\ifpreprintsty\else\end{multicols}\fi}
%
%
%\draft
%\widetext
%\begin{document}
\title{Understanding amorphous phase-change materials from the viewpoint of Maxwell rigidity}
\author{M. Micoulaut$^1$, J.-Y. Raty$^2$ , C. Otjacques$^2$  and C. Bichara$^3$}
\affiliation{$^1$ Laboratoire de Physique Théorique de la Matière Condensée,
Université Pierre et Marie Curie,  Boite 121, 4, Place Jussieu, 75252
Paris Cedex 05, France\\
$^2$ Physique de la Matière Condensée, B5, Université de Liège
B4000 Sart-Tilman, Belgium\\
$^3$ Centre Interdisciplinaire de Nanoscience de Marseille, CNRS et Universités d'Aix-Marseille, Campus de Luminy, Case 913, 13288 Marseille, France}

\date{\today}
%\maketitle
\begin{abstract}
Phase-change materials (PCMs) are the subject of considerable interest because they have been recognized as potential active layers for next-generation non-volatile memory devices, known as Phase Change Random Access Memories (PRAMs). By analyzing First Principles Molecular Dynamics simulations we develop a new method for the enumeration of mechanical constraints in the amorphous phase and show that the phase diagram of the most popular
system (Ge-Sb-Te) can be split into two compositional regions having a well-defined mechanical character: a Tellurium rich flexible phase, and a stressed rigid phase that encompasses the known PCMs. This sound atomic scale insight should open new avenues for the understanding of PCMs and other complex amorphous materials from the viewpoint of rigidity.
\end{abstract}
\maketitle
%\bcols
Driven by applications in data storage \cite{PCM} fundamental and applied studies of tellurides are rapidly developing. The most promising phase-change materials (PCMs) belong to the ternary Ge-Sb-Te system with particular compositions such as the Ge$_2$Sb$_2$Te$_5$ already used in industrial products \cite{storage}. To optimize the peculiar property portfolio of these PCMs, a key issue is the understanding of their atomic structure. This has led to a series of investigations of the structure of both amorphous and crystalline phases using experimental as well as computer simulation techniques (for a review, see \cite{PCM}).
\par
Since they are related to the ageing of PCMs \cite{IP}, the mechanical properties of the amorphous phase are of major interest. In parent systems where the 8-N rule holds (N: number of s and p electrons), particularly sulphur and selenium based amorphous networks, rigidity theory offers a practical computational scheme using topology, namely the Maxwell counting procedure, and has been central to many contemporary investigations on non-crystalline solids \cite{IP, rigidity}. It has led to the recognition of a rigidity transition \cite{JCP79} which separates flexible glasses, having internal degrees of freedom that allow for local deformations, from stressed rigid glasses which are "locked" by their high bond connectivity.
\par
What happens if these elements are replaced by the heavier element Te which will lead to more complicated local structures, as highlighted both from experiments \cite{Kolobov, Paessler} and simulations \cite{Akola, Caravati, Bichara2007}? Does the counting procedure still hold? Attempts in this direction have been made on a heuristic basis \cite{Paessler} but they seem to contrast with experimental observations. A firm basis for the Maxwell constraint counting is therefore very much desirable to assess algorithms specially designed for PCMs. This is the purpose of the present study that develops a precise enumeration algorithm for constraints arising from bond-stretching (BS) and bond-bending (BB) interactions, based on the analysis of atomic scale trajectories using First Principles Molecular Dynamics Simulations (FPMD). Combined with rigidity theory, it opens an interesting perspective to study amorphous phase change materials in much the same fashion as network glasses. As a result, we show that the phase diagram of the Ge-Sb-Te system can be separated into two compositional regions having a well-defined mechanical character derived from rigidity theory: a flexible Te-rich phase, and a (Sb,Ge)-rich phase that is stressed rigid. The most commonly used GST phase change materials belong to this second category.
\par
	At the heart of the rigidity concept is the identification of relevant interatomic forces between atoms in a manner similar to what Maxwell pioneered for trusses and macroscopic structures \cite{Maxwell}. When applied to covalent amorphous networks and once the forces acting as constraints are identified (BS and BB forces), a similar analysis can be performed leading to the Phillips-Thorpe rigidity transition \cite{JCP79}, which separates flexible (underconstrained) networks from stressed rigid (overconstrained) networks \cite{rigidity,IP}. As in standard mechanics however, instead of treating forces and querying about motion, one can ask the opposite question and try to relate motion to the absence of a restoring force. Using FPMD, we generate therefore atomic scale trajectories of various amorphous systems at low temperature using an electronic structure model (see \cite{gese2, Raty_FPMD} and EPAPS supplementary material for simulation details) and apply a structural analysis in relation with rigidity theory. The number of neighbors, and hence the number of BS constraints, is calculated by integrating the radial distribution functions up to its first minimum (Table I). To estimate the number of bond-bending constraints we analyze the partial bond angle distributions. For each type of central atom $0$, the six first neighbors $i$ are selected and sorted according to their distances, as done in \cite{Bichara96}, and the distributions  P($\theta_{ij}$) of the 15 corresponding angles $i0j$ ($i$=1..5, $j$=2..6) are calculated, i.e. 102, 103, 203, etc. The second moment $\sigma_{\theta_{ij}}$ of P($\theta_{ij}$) provides a quantitative estimate of the angular excursion around the mean value of angle $i0j$, thus measuring the strength of the bond-bending restoring force. An angle displaying a wide $\sigma_{\theta_{ij}}$ corresponds to a broken BB constraint as there is a weak interaction to maintain the angle fixed. In an opposite way, sharp bond angle distributions lead to intact constraints.
\par
\begin{figure}
\begin{center}
\epsfig{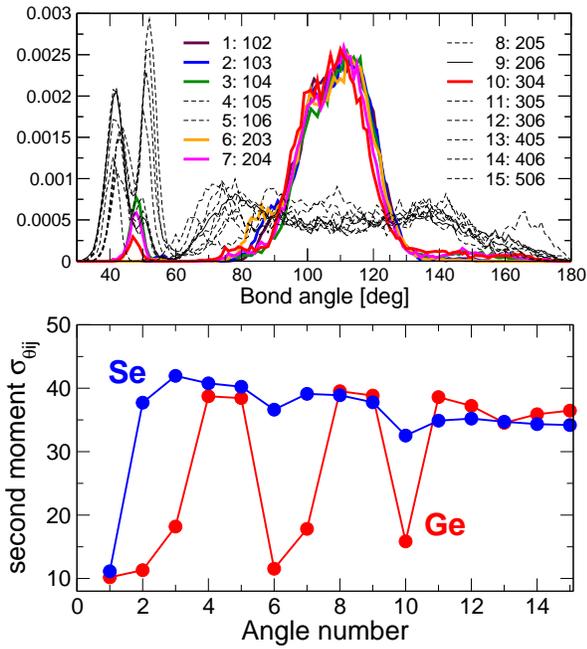}
\end{center}
\caption{(color online) a) Ge-centered bond angle distributions (up to 15) for various triplets of atoms $i0j$ ($i$=1..6, $j$=2..5) in amorphous GeSe$_2$.
The six colored distributions have a low second moment (typically 10-20$^o$, see panel b). The angle number assignments (1..15) displayed in the Ge panel
are valid for all other angular studies. b) Second moment $\sigma_{\theta_{ij}}$ of the distributions as a function of the angle number
in amorphous GeSe$_2$: Ge-(red) and  Se-centered angles (blue).}
\end{figure}
In order to check this method, we first apply it to the benchmark case GeSe$_2$ , for which application of constraint counting algorithms is straightforward \cite{JCP79}. According to the Phillips-Thorpe enumeration, one has for a r-coordinated atom respectively r/2 and 2r-3 BS and BB constraints. Thus a four-fold Ge atom has 2 BS and 5 BB constraints whereas the two-fold selenium atom has 1 BS and 1 BB constraint leading on the overall to $n_c=3.67$ constraints per atom \cite{JCP79}. We obtain the coordination numbers $r_{Ge}$=4.04 and $r_{Se}$=1.98 from the area of the first peak of
the Ge- and Se- centered pair distribution functions. We furthermore find that $\sigma_{\theta_{ij}}$ can vary between 10$^o$ and 40$^o$ depending on the different angles $i0j$ considered (Fig. 1). For the Ge-centered atoms, six moments $\sigma_{\theta_{ij}}$ are found to be of the order of 10-20$^o$, very well separated from all others for which $\sigma_{\theta_{ij}}$ $\simeq$ 40$^o$.
However there is one redundant constraint that needs to be removed because it can be determined from the five other angles. This leaves the estimate with 5 independent BB constraints for the Ge atom. For the Se atom, a single low $\sigma_{\theta_{ij}}$  (i.e. a single BB constraint) is found ($12^o$) around the mean value $\bar \theta_{ij}=100^o$, in agreement with experiment
\cite{Rennes}. We arrive to the conclusion that the constraint computation from FPMD matches exactly the direct counting from \cite{JCP79}.
\par
\begin{figure}
\begin{center}
\epsfig{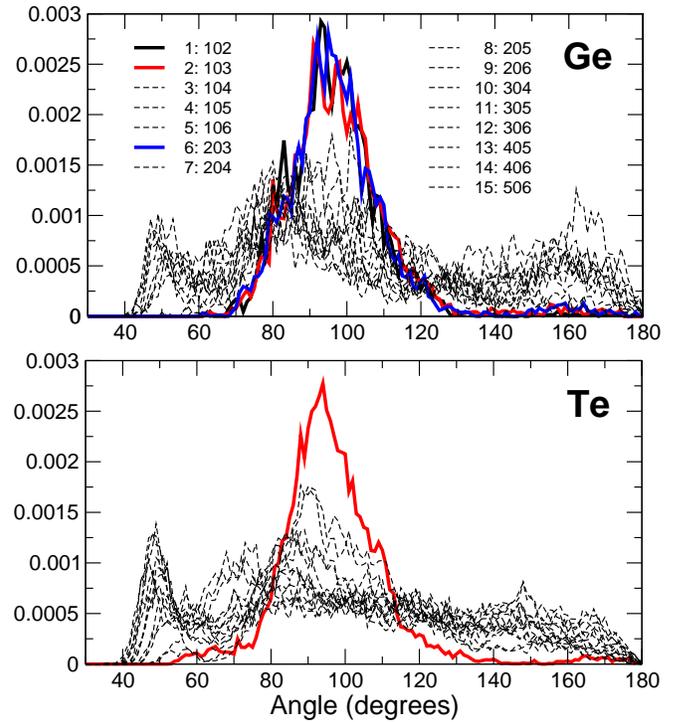}
\end{center}
\caption{(color online) Ge and Te centered bond angle distribution in the Ge$_1$Sb$_2$Te$_4$ (124) amorphous system.
The curves in color correspond to distributions with a low second-moment $\sigma_{\theta_{ij}}$ considered as having intact BB
constraints. The peaks around 50$^o$ correspond to the steric hindrance (hard core repulsion between neighboring atoms).}
\end{figure}
	Having validated the method with GeSe$_2$, we now turn to the amorphous Ge-Sb-Te system and focus on seven particular compositions, namely Ge$_1$Sb$_2$Te$_4$ (124), Ge$_2$Sb$_2$Te$_5$ (225), GeTe (101), GeTe$_6$ (106), GeSb$_6$ (160), Sb$_2$Te (021) and Sb$_2$Te$_3$ (023) (see also Fig. 4), using extensive FPMD simulations. We determine the BS constraints from the coordination numbers extracted from the partials (Table I). The coordination number of Ge and Sb is nearly equal to r=4, with a preference for heteropolar bonding with Te atoms, which have a coordination number between 2.1 and 2.9, larger than the 8-N value (r=2).
\par
\begin{table}
\begin{center}
\begin{tabular}{llccccccccccc}\hline
Compound& & &Atom& & &$r_i$& & &$n_i^{BB}$& & &$n_c$\\ \hline\hline
 & & & & & & & & & & & & \\
GeTe$_6$& & &Ge& & &4.0& & &3.3& & &    \\
        & & &Te& & &2.4& & &1.0& & &2.68\\ \hline
GeTe    & & &Ge& & &4.1& & &3.0& & &    \\
        & & &Te& & &2.9& & &1.0& & &3.75\\ \hline
Ge$_1$Sb$_2$Te$_4$& & &Ge& & &4.0& & &3.4& & & \\
                  & & &Sb& & &4.1& & &3.0& & & \\
                  & & &Te& & &2.8& & &1.0& & &3.59\\ \hline
Ge$_2$Sb$_2$Te$_5$& & &Ge& & &4.0& & &3.2& & & \\
                  & & &Sb& & &3.8& & &3.0& & & \\
                  & & &Te& & &2.4& & &1.0& & &3.47 \\ \hline
GeSb$_6$& & &Ge& & &4.1& & &5.0& & & \\
        & & &Sb& & &3.7& & &3.0& & &5.16 \\ \hline
Sb$_2$Te& & &Sb& & &4.0& & &3.0& & & \\
        & & &Te& & &2.5& & &1.0& & &4.08 \\ \hline
Sb$_2$Te$_3$& & &Sb& & &3.7& & &3.0& & & \\
            & & &Te& & &2.1& & &1.0& & &3.17 \\ \hline

\end{tabular}
\end{center}
\caption{Coordination number $r_i$ of the atomic species, giving the number of bond-stretching (BS) constraints ($r_i$/2),  number of BB constraints $n_i^{BB}$ computed from the second moments of the bond angle distributions $P_(\theta_{ij})$, and total number of constraints $n_c$ in the seven different Ge-Sb-Te compounds.}
\end{table}

The Ge and Te centered P($\theta_{ij}$) for the 124 compound are displayed in Fig. 2 . Certain angles clearly display a limited motion around their mean value. Similar figures are found for Sb (not shown) from which the appropriate counting can be drawn. Fig. 3 shows the 15 different second moments $\sigma_{\theta_{ij}}$ for the compositions 124 and 225 leading to the determination of corresponding BB constraints for Ge, Sb and Te atoms. Compared to the benchmark system GeSe$_2$, we notice that $\sigma_{\theta_{ij}}$ is more scattered for large angle number n (i.e. n$>$6), which suggests an increased orientational disorder when more distant neighbors are considered.
\par
In the ternary compositions 124 and 225, only three second moments are of the order of $\sigma_{\theta_{ij}}\simeq$ 10-15$^o$ for the Ge and Sb atoms, associated with well-defined angles at $\bar \theta_{ij}$=90-100$^o$ which are, together with those found at $\simeq$180$^o$, reminiscent of the distorted octahedral-like rocksalt cubic phase \cite{Kolobov, Welnic}. The present
results contrast with the view that would follow the standard enumeration of constraints, directly derived from coordination numbers obeying the 8-N rule. In fact, a three-fold Sb would give rise to 1.5 BS and 3 BB constraints \cite{JCP79}. Here, Sb has an additional neighbor that increases the number of BS constraints but it does not give rise to two additional BB constraints (Fig 3).
\par
Although it is found $r_{Te}>2$, Te has only one angular constraint (Fig. 3, $\sigma_{\theta_{ij}}$=12$^o$ in the 124 and $\bar \theta_{ij}\simeq $95$^o$), the two other possible angles (angle number 2: 103 and 6: 203) being much more
flexible ($\sigma_{\theta ij}$=27$^o$ and 29$^o$). On the basis of this enumeration, and using results of Table I and of Fig. 3, a Maxwell estimate for the number of BS and BB constraints of Ge$_x$Sb$_y$Te$_{1-x-y}$ is given by :
\begin{eqnarray}
\label{1}
n_c&=&{\frac {1}{2}}\biggl[x(r_{Ge}-r_{Te})+y(r_{Sb}-r_{Te})+r_{Te}\biggr]\\ \nonumber
&+&\biggl[x(n_{Ge}^{BB}-n_{Te}^{BB})+y(n_{Sb}^{BB}-n_{Te}^{BB})+n_{Te}^{BB}\biggr]
\end{eqnarray}
where the square brackets are used to separate BS from BB contributions. Results for the seven compositions are given in Table I. They furthermore take into account the possibility of two local environments for four-fold Ge in the presence of Te also found in \cite{Akola, Caravati}: a majority of distorted octahedral sites having 3 constraints for the angles $\bar \theta_{ij}$=90-100$^o$ (see Fig. 3), and a minority of tetrahedral Ge (calculated to have a respective fraction of $\eta=$0.1 and $\eta$=0.2 in the 225 and 124) which have 5 BB constraints as in GeSe$_2$. This means that the average number of Ge BB constraints is $n_{Ge}^{BB}=5\eta+3(1-\eta)=3+2\eta$ and leads finally to $n_c^{124}$=3.59 and $n_c^{225}$=3.47, $r_{Te}$ being calculated for each composition using Table I. One can thus conclude that 124 and 225 are stressed rigid, i.e. they have more constraints than degrees of freedom (3 in 3D). The present results contrast with a constraint enumeration based on EXAFS measurements, and with the assumption that GST materials are perfect glasses \cite{Paessler}, but they agree with the obvious observation that, apart the 106 alloy \cite{GeTe}, which is found flexible, but close to the optimal $n_c=3$, none of the alloys studied can form bulk glasses.
\begin{figure}
\begin{center}
\epsfig{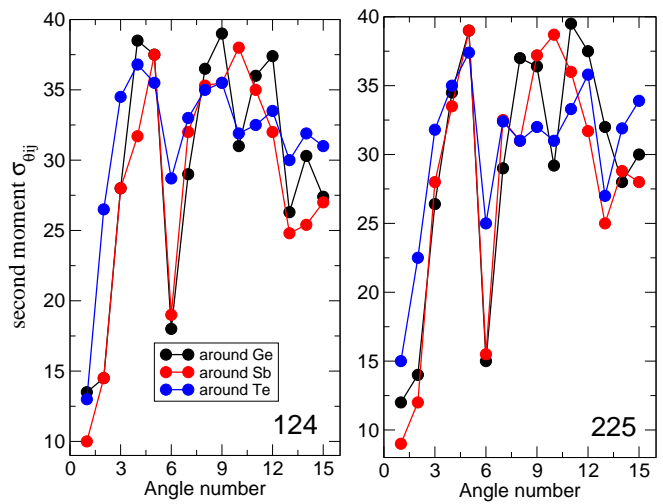}
\end{center}
\caption{(color online) Angular Second-moment $\sigma_{\theta_{ij}}$ of the bond angle distributions of 124 and 225 amorphous systems as a function of angle number for Ge-(black), Sb-(red) and Te-centered (blue) atoms. The correspondence between angle number (x-axis) and angles (i0j) with 0 the central atom and $i$ and $j$ the neighbors, is given in Fig. 1 and 2. Only three Ge and Sb (one Te) angles have a low $\sigma_{\theta_{ij}}$, corresponding to intact bond-bending constraints.}
\end{figure}
Using these elements, we now determine an approximate flexible to rigid transition composition \cite{JCP79} from the Maxwell estimate  corresponding to $n_c=3$. Considering the numbers given in Table I and the results of Fig. 3, we make the simple assumption that Ge has four neighbors ($r_{Ge}$=4) with a fraction $\eta$ of tetrahedral sites, Sb has four neighbors ($r_{Sb}$=4) in a distorted octahedral geometry ($n_{Sb}^{BB}$=3) and Te has $r_{Te}$=2.6
neighbors and one BB constraint. Pure amorphous Ge is known to be a tetrahedral network and in GeSb$_6$ all Ge are tetrahedral, so that $\eta$=1. From table I, we note that the addition of Te effectively lowers the fraction $\eta$ of
tetrahedral Ge, so that, for the sake of simplicity, we assume that $\eta=x+y$, in which case one finally obtains a parameter-free rigidity transition line depending only on the compositions $x$ and $y$:
\begin{eqnarray}
\label{eq2}
y={\frac {7}{27+20x}}-x
\end{eqnarray}
The relationship (\ref{eq2}) is found to be close to the compositional join GeTe$_4$-SbTe$_4$ (Fig. 4), and defines two regions in the GST triangle. In the Te-rich region, the system has not enough Ge or Sb cross-links to ensure
rigidity, and local deformations are allowed. In the second region, where usual PCMs are found, the amorphous phases are stressed rigid. Bulk glass formation seems to be only possible in the flexible phase as shown from experimental
data \cite{Lebaudy}.
\par
\begin{figure}
\begin{center}
\epsfig{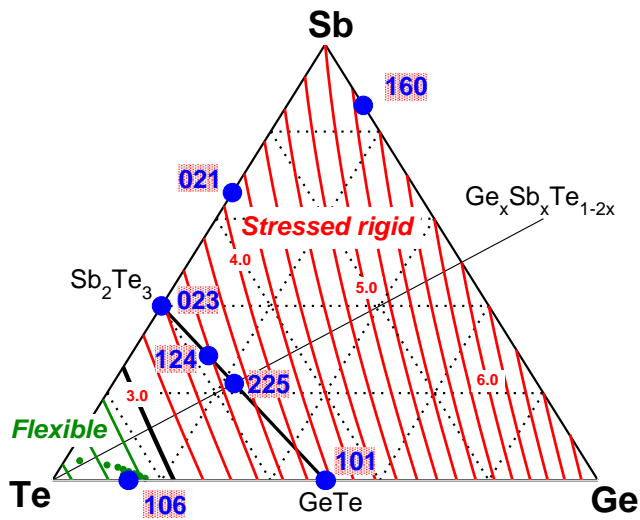}
\end{center}
\caption{(color online) Contour map of the number of constraints $n_c$ in the ternary Ge$_x$Sb$_y$Te$_{1-x-y}$ phase
diagram. The red and green lines correspond respectively to the stressed rigid and flexible phase. Blue circles represent
the compositions studied by FPMD in this Letter. The thick black line represents the rigidity transition line defined
by equ. (2), and separates the flexible (Te-rich) from the stressed rigid phase where most PCMs can be found, especially
on the GeTe-Sb$_2$Te$_3$ tie line (black line). Green dots in the flexible phase represent bulk glass compositions
obtained experimentally \cite{Lebaudy}.}
\end{figure}
\par
In summary, we have developed a new constraint counting algorithm applicable to tellurides for which a simple counting based on the 8-N rule does not apply in a straightforward manner. We show that atomic-scale trajectories obtained
from First Principles Molecular Dynamics simulations can be appropriately used for the estimation of bond-stretching and bond-bending constraint counting and applied to the GST phase-change system. The results show that amorphous systems lying on the popular Sb$_2$Te$_3$-GeTe tie-line in the GST compositional triangle belong to a stressed rigid phase, whereas an rigidity transition line is obtained close to the SbTe$_4$-GeTe$_4$ join. Furthermore, since an intermediate phase with some remarkable properties (absence of ageing, stress-free character and space-filling tendencies) \cite{IP} has been found close to $n_c=3$ in sulphide and selenide systems, one may wonder to what extent these properties can be observed in tellurides as well, and how these properties, once being observed, could be used in close future to design phase-change materials with the corresponding functionality.

%\ecols

\begin{thebibliography}{100}

\bibitem{PCM} {\em Phase Change Materials: Science And Applications}, S. Raoux, M. Wuttig Eds., (Springer, 2008).

\bibitem{storage}   H-Y. Cheng, C.A. Jong, R.-J. Chung, T.-S. Chin, R.-T. Huang, Semicond. Sci. Technol. {\bf 20}, 1111 (2005).

\bibitem{IP} {\em Rigidity and Boolchand phases in nanomaterials}, M. Micoulaut, M. Popescu, Eds. (INOE Publishing House Bucarest, 2009).

\bibitem{rigidity} {\em Rigidity theory and applications}, M.F. Thorpe and P.M Duxbury Eds.  (Kluwer Academic, Plenum Publishers New York, 1999).

\bibitem{JCP79} J.C. Phillips, J. Non-Cryst. Solids {\bf 34}, 153 (1979); M.F. Thorpe, J. Non-Cryst. Solids {\bf 57}, 355 (1983).

\bibitem{Kolobov} A. V. Kolobov, P. Fons, A. I. Frenkel, A. L. Ankudinov, J. Tominaga,  T. Uruga, Nature Mat. {\bf 3}, 703 (2004).

\bibitem{Paessler} D. A. Baker, M.A. Paessler, G. Lucovsky, S.C. Agarwal, P.C. Taylor,  Phys. Rev. Lett. {\bf 96}, 255501 (2006)

\bibitem{Akola} J. Akola, R.O. Jones, Phys. Rev. B {\bf 76}, 235201 (2007).

\bibitem{Caravati} S. Caravati, M. Bernasconi, T.D. Kuehne, M. Krack, and M. Parrinello, Appl. Phys. Lett. {\bf 91} , 171906 (2007).

\bibitem{Bichara2007} C. Bichara, M. Johnson, and J.-P. Gaspard, Phys. Rev. B {\bf 75}, 060201 (2007).

\bibitem{Maxwell}  J.C. Maxwell, Phil. Mag. {\bf 27}, 294 (1864).

\bibitem{gese2} M. Micoulaut, R. Vuilleumier, C. Massobrio, Phys. Rev. B {\bf 79}, 214204 (2009).

\bibitem{Raty_FPMD} R. Shaltaf, E. Durgun, J.-Y. Raty, Ph. Ghosez, X. Gonze, Phys. Rev. B {\bf 78}, 205203 (2008); J.Y. Raty, C. Otjacques, J.P. Gaspard, C. Bichara, Solid State Sciences in press (2009).

\bibitem{Bichara96} C. Bichara, J.Y. Raty and J.-P. Gaspard, Phys. Rev. B {\bf     53}, 206-11 (1996).
    
\bibitem{Rennes} B. Bureau, J. Troles, M. Le Floch, P. Guénot, F. Smektala, J. Lucas, J. Non-Cryst. Solids {\bf 319}, 145 (2003).

\bibitem{Welnic} W. Welnic, A. Pamungkas, R. Detemple, C. Steimer, S. Blügel, M. Wuttig, Nature {\bf 5}, 56 (2006).

\bibitem{GeTe} D. Selvanathan, R.N. Enzweiler, W. J. Bresser, P. Boolchand, Bull. Am. Phys. Soc. {\bf 42}, 249 (1997).

\bibitem{Lebaudy} P. Lebaudy, J.M. Saiter,  J. Grenet,  M. Belhadji, and C. Vautier, Materials Sci. Eng. A {\bf 132}, 132 (1991).

\end{thebibliography}
\end{document}